\documentclass[12pt,english]{article}

\usepackage{jheppub}

\usepackage{amsmath,amssymb,array,mathrsfs,amsfonts}
\usepackage{epsfig}
\usepackage{euscript}

\usepackage{graphicx}
\newcommand{\lyxdot}{.}

\makeatletter

\title{Complex Langevin Dynamics for chiral Random Matrix Theory}
\author[a]{A. Mollgaard}
\author[a]{K. Splittorff}
\affiliation[a]{Discovery Centre, Niels Bohr Institute, University of Copenhagen, Blegdamsvej 17, 2100 Copenhagen {\O}, Denmark}
\abstract{
We apply complex Langevin dynamics to chiral random matrix theory  
at nonzero chemical potential. At large quark mass the simulations 
agree with the analytical results while incorrect convergence is 
found for small quark masses. The region of quark masses for which 
the complex Langevin dynamics converges incorrectly is identified as 
the region where the fermion determinant frequently traces out a 
path surrounding the origin of the complex plane during 
the Langevin flow. This links the incorrect convergence to an 
ambiguity in the Langevin force due to the presence of the logarithm 
of the fermion determinant in the action.}
\notoc
\begin{document}

\maketitle

\newpage

\section{Introduction}

Understanding the phase diagram of QCD in the chemical potential, $\mu$, 
and temperature, $T$, plane
is one of the greatest challenges of high energy physics today. At
$\mu=0$, QCD may be treated on the lattice, but for $\mu\neq0$ the
fermion determinant turns complex
\begin{equation}
\det\equiv\det\left(i\gamma_{\nu}D_{\nu}+\mu\gamma_{0}+m\right)\in\mathbb{C},
\end{equation}
such that a reweighting is needed to apply importance sampling. The
numerical computations grow exponentially harder as the volume is increased,
and this complication has been named the ``the sign problem'' (see
\cite{Aarts:2013bla,deForcrand:2010ys,Splittorff:2006vj} for reviews). 

An interesting method for generating field configurations according
to the probabilistic weight at $\mu=0$ is that of Langevin dynamics 
\cite{Damgaard:1987rr}.
Unlike other approaches, Langevin dynamics may naturally be generalized
to the case of complex weights, but tools to check the results of
such simulations have not been available until recently \cite{AartsFirst,AartsSecond,AartsThird}.
Not only have we gained improved understanding; complex Langevin dynamics
has also been applied to solve sign problems in a number of physical
models such as the relativistic Bose gas \cite{Aarts:2008wh,Aarts:2009hn,Aarts:2009yj}
and a one-dimensional version of QCD \cite{Aarts:2010gr}. Moreover, 
the technique
of gauge cooling has very recently been introduced to obtain well
behaved dynamics in the case of QCD in the heavy quark mass limit
\cite{Seiler:2012wz,Aarts:2013uxa} as well as in full QCD 
\cite{Sexty:2013ica}. In this paper we apply complex Langevin dynamics to 
chiral random matrix theory at nonzero chemical potential 
\cite{Osborn:2004rf,Bloch:2012bh}. Just as in QCD the sign problem 
of chiral random matrix theory comes in through the
presence of a complex valued fermion determinant. The advantage of the 
random matrix theory is that exact analytic solutions are known and 
allow for a direct test of the complex Langevin simulation.

Our main objective with the simulation of chiral random matrix theory 
is to understand to which degree complex Langevin dynamics (CLD) 
is able to deal 
with a complex valued fermion determinant. The standard approach of 
complex Langevin is first to exponentiate the fermion determinant,
\begin{equation}
\det(M) \, e^{-S_{g}}=e^{-S_{g}+\log\left(\det(M)\right)},
\end{equation}
and subsequently obtain the Langevin drift term by taking the 
derivative of the action, $S\equiv S_{g}-\log\left(\det(M)\right)$, with 
respect to the relevant fields. 
At this point, however, one is faced with an ambiguity because the 
logarithm is a multivalued function. On one hand 
one could argue that the derivative of the logarithm in the complex 
plane is well defined on the infinite Riemann sheet and simply 
extend the standard derivative, $d\log f(x)/dx = f'(x)/f(x)$, into the full 
complex plane. On the other hand one could choose to work with 
principal part of the logarithm which is single-valued, but has a branch 
cut from the origin along 
the negative real axis. The imaginary part is discontinuous across this cut and the derivative is therefore not straightforward 
to implement in the Langevin force. However, since the imaginary part 
of the logarithm is exactly what allows $\exp(\log\left(\det(M)\right))$
to change sign when the determinant moves from positive to negative
values on the real axis it is natural to expect that the influence 
of the cut on the Langevin flow is essential.

In the Langevin simulation of the chiral random matrix theory we 
will follow \cite{Aarts:2013uxa,Seiler:2012wz,Sexty:2013ica} and 
make use of the standard form of the derivative of the logarithm.  
We observe that while the Langevin dynamics reproduce the analytical 
results at large values of the quark mass, failed convergence is 
found for small masses. We identify the range of quark masses for which 
the complex Langevin dynamics converges to incorrect values as the region 
where the determinant frequently traces out a path surrounding the origin  
of the complex plane. The results of the Langevin dynamics in this region  
are similar in nature to the results obtained in the phase quenched chiral 
random matrix theory. This suggests that the cut is relevant for 
the Langevin dynamics and should not be ignored in the flow equations.

As the derivative of the logarithm is unique on the full principal 
branch the 
ambiguity of the Langevin flow is only relevant when the determinant 
frequently circles the origin. We use this to predict regions of 
successful and failed convergence in two previously studied $U(1)$ models 
\cite{Aarts:2008rr,Aarts:2010gr}. It is likely also to explain the 
crossover from successful to failed complex Langevin dynamics 
in other models with a logarithm in the action such as the Thirring model 
\cite{Pawlowski:2013pje}.

The outline of the article is the following: in section 2 we give
a more detailed introduction to the Langevin equation, and in section
3 we take a closer look at the logarithm. Section 4 is devoted to
the application of complex Langevin dynamics to chiral random matrix 
theory, and the 
two $U\left(1\right)$ models are discussed in section 5. 
Finally in section 6 we summarize the results and offer a look ahead.
In the appendix we revisit the $U(1)$ model of \cite{Ambjorn:1986fz}.

\section{The Langevin equation}

The real Langevin equation provides a method for generating ensembles of
field configurations according to the positive weight function at
$\mu=0$. For simplicity let us first consider a single real 
variable, $x$, with a partition function 
\begin{equation}
Z=\int e^{-S\left(x\right)}dx,
\end{equation}
and a positive weight 
\begin{equation}
e^{-S\left(x\right)}\in\mathbb{R}_{+}.
\end{equation}
The real Langevin equation then takes the form
\begin{equation}
x(t+dt)=x(t)-\partial_{x}S\left(x\left(t\right)\right)\cdot dt+dW,\label{eq: Langevin equation}
\end{equation}
where $dW$ is a stochastic variable of zero mean $\left\langle dW\right\rangle =0$
and variance $\left\langle dW^{2}\right\rangle =2dt$. Analytic mean
values
\begin{equation}
\left\langle O\left(x\right)\right\rangle =\frac{\int dx\, O\left(x\right)e^{-S\left(x\right)}}{Z}\label{eq: analytical mean}
\end{equation}
may be calculated numerically by updating the variable according to
(\ref{eq: Langevin equation}) and calculating the observable value
with each step; the mean value of such measurements approaches the
analytic result as $t\rightarrow\infty$ \cite{SPP:bog}. 

As proposed by Parisi \cite{Parisi} and Klauder \cite{Klauder:1983nn}
we may generalize the Langevin equation to complex weights using the 
drift 
\begin{equation}
\frac{dS(x)}{dx}|_{x\rightarrow x+iy}.
\label{drift}
\end{equation}
The variable is then pushed off the real axis and into the complex
plane, $x\rightarrow z=x+iy$, where Langevin measurements may be
performed as in the case of real and positive weights. The proof relating
Langevin dynamics to the path integral no longer applies though (see
\cite{Aarts:2013uxa}), and simulations only converge correctly some
of the time. In order to have a well defined drift term, (\ref{drift}), 
in the whole complex plane, we need the action to be entire, such 
that the Langevin equation may be written uniquely as 
\begin{equation}
z(t+dt)=z(t)-\partial_{z}S\left(z\left(t\right)\right)\cdot dt+dW.
\label{Leq}
\end{equation}
Two criteria for correct convergence have been given for such entire
actions in \cite{AartsFirst,AartsSecond,AartsThird}.

The Langevin equation trivially generalizes to theories with more 
degrees of freedom, such as QCD on the lattice. The Euclidian
partition function is given as an integral over the link variables
$U$ and has the structure 
\begin{equation}
Z=\int\mathcal{D}U\,\det M\cdot e^{-S_{g}},
\end{equation}
where $S_{g}$ is the Yang-Mills term and $\det M$ is the fermion
determinant (see eg \cite{LG}). For $\mu\neq0$ the fermion
determinant turns complex, such that for example a reweighting must 
be performed or the complex Langevin equation must be applied. Reweighting is 
exponentially hard in the volume \cite{Splittorff:2006fu} because of the sign problem, so we turn
to complex Langevin dynamics (CLD). In order to go beyond the quenched 
approximation we include the fermion determinant
into the action
\begin{equation}
S=S_{g}-\log\left(\det M\right),\label{eq: QCD effective}
\end{equation}
which determines the Langevin flow through (\ref{Leq}). 
As we now discuss, such a logarithmic term in the action leads to an 
ambiguity in the complex Langevin dynamics.

\section{The logarithm}

The logarithm is a multivalued function with output
\begin{equation}
\log(z)=\mbox{Log}(z)+i2\pi\cdot n,\label{eq: logarithm}
\end{equation}
where ${\rm Log}(z)$ is the principal part of the logarithm and $n\in\mathbb{Z}$. 
 Any definition of the logarithm compatible with (\ref{eq: logarithm})
yields the same weight in the original path integral, but the Langevin
flow is determined from the action alone, so here it matters what
definition is chosen. If we accept the use of a multivalued logarithm,
then the derivative of the logarithm may be written as
\begin{equation}
\partial_{z}\log\left(z\right)=1/z.
\end{equation}
If we require a single valued logarithm and work with the principal part 
of the logarithm, then the derivative on the branch is still
\begin{equation}
\partial_{z}\mbox{Log}\left(z\right)=
1/z, \: z\in\mathbb{\mathbb{C}}/\mathbb{R}_{-}.
\end{equation}
In addition, however, there will be a cut from the origin out to infinity, 
where the derivative is singular, see eg.~\cite{FZ} for a discussion. 

In the next section we will adopt the first option where the cut is ignored, 
and demonstrate that simulations of chiral random matrix theory yield 
failed measurements, when the phase of the determinant, $\det M$, frequently 
circles the full range $\left[-\pi,\pi\right]$ during the Langevin simulation.

\section{chiral Random Matrix Theory}

Instead of approaching QCD head on we study chiral random matrix theory
\cite{chRMT1,chRMT2,chRMT3} with nonzero chemical potential, which has 
a similar structure with a 
fermion determinant in the measure, but at the same time is much
simpler and possible to solve analytically. Chiral random matrix theory
has already provided several deep insights into QCD at nonzero chemical 
potential: it has explained the failure of the quenched approximation 
\cite{misha}, it has uncovered the OSV relation \cite{OSV} which replaces 
the Banks-Casher relation \cite{BC} at nonzero chemical potential and 
it has revealed the surprising phase structure of QCD with bosonic 
quarks at nonzero chemical potential \cite{Splittorff:2006uu}. 
The reason why the far 
simpler random matrix theory can give direct insights into QCD is that the
quark mass dependence of the chiral condensate and baryon density are
uniquely determined by the flavor symmetries in the microscopic limit. 
Since QCD and chiral random matrix theory have the exact same flavor 
symmetries, we can use the analytic tools of chiral random matrix theory
to derive the universal predictions for QCD.

The partition function reads
\begin{eqnarray}
Z_{N}^{N_{f}}(m) & = & \int d\Phi d\Psi\:\mbox{det}^{N_{f}}\left(D(\mu)+m\right)\exp\left(-N\cdot\mbox{Tr}[\Psi^{\dagger}\Psi+\Phi^{\dagger}\Phi]\right),\label{eq: RMT model 1}
\end{eqnarray}
where
\begin{equation}
D(\mu)+m=\left(\begin{array}{cc}
m & i\cosh(\mu)\Phi+\sinh(\mu)\Psi\\
i\cosh(\mu)\Phi^{\dagger}+\sinh(\mu)\Psi^{\dagger} & m
\end{array}\right).\label{eq: D+m}
\end{equation}
The degrees of freedom, $\Psi$ and $\Phi$, are general complex
$N\times N$ matrices, so $\Phi_{ij}=a_{ij}+ib_{ij}$, $\Psi_{ij}=\alpha_{ij}+i\beta_{ij}$
and $d\Phi d\Psi=dadbd\alpha d\beta$. This chiral random matrix 
theory was introduced in \cite{Osborn:2004rf}, but uses the redefined 
parameters of \cite{Bloch:2012bh}. Notice that 
\begin{equation}
\mbox{det}^{\ast}(D\left(\mu\right)+m)=\mbox{det}(D\left(-\mu^{\ast}\right)+m)\label{eq: det*}
\end{equation}
just as in QCD. In the microscopic limit $N\rightarrow\infty$ with 
$\tilde{m}=Nm$ and $\tilde{\mu}=\sqrt{N}\mu$ kept constant, the matrix model may
be shown to be equivalent to the $\epsilon$-regime of chiral perturbation
theory \cite{Osborn:2004rf,Akemann:2004dr,Basile:2007ki,Akemann:2007rf}.

\begin{figure}
\begin{centering}
\includegraphics[width=10cm]{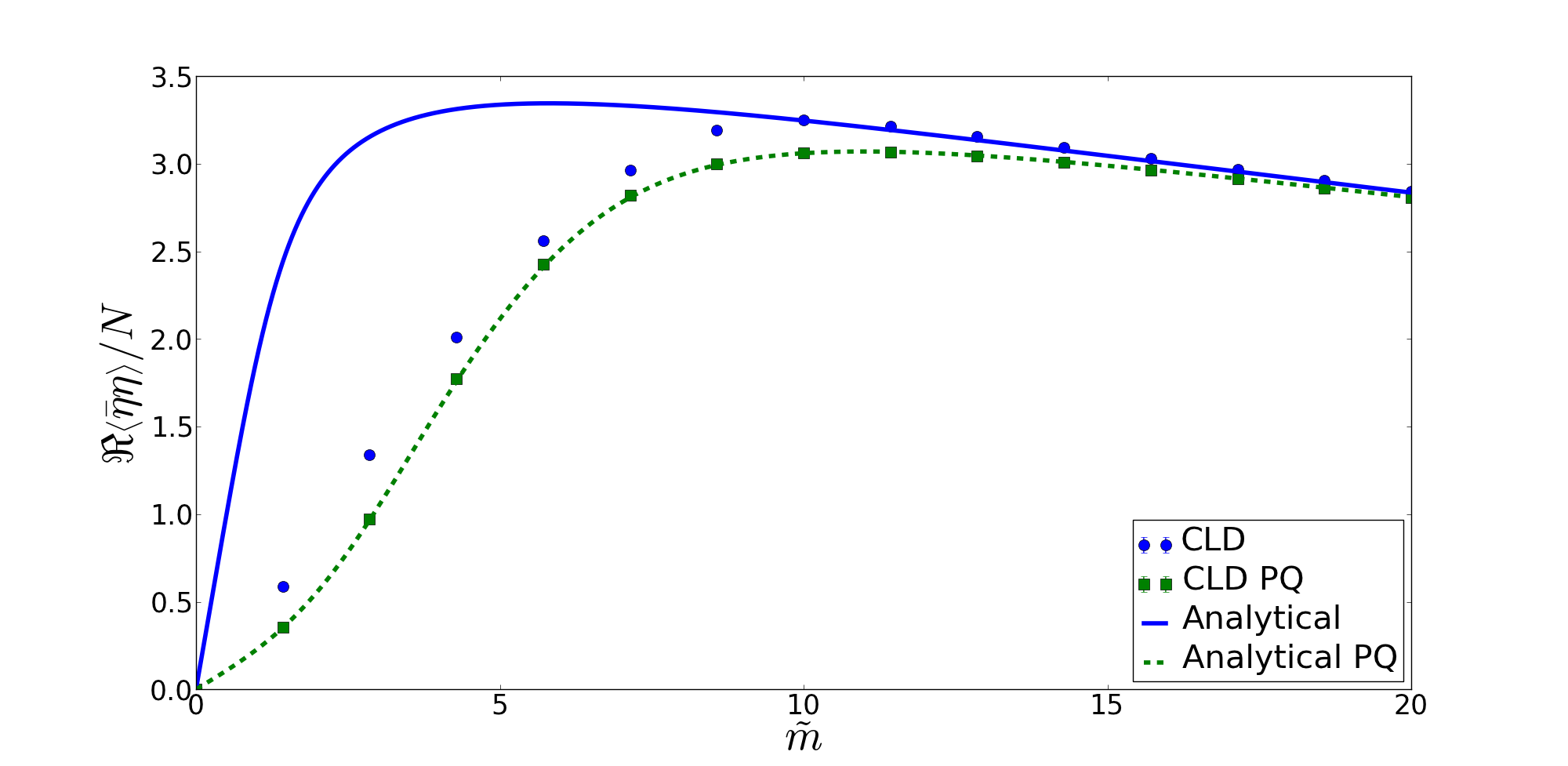}
\par\end{centering}

\begin{centering}
\includegraphics[width=10cm]{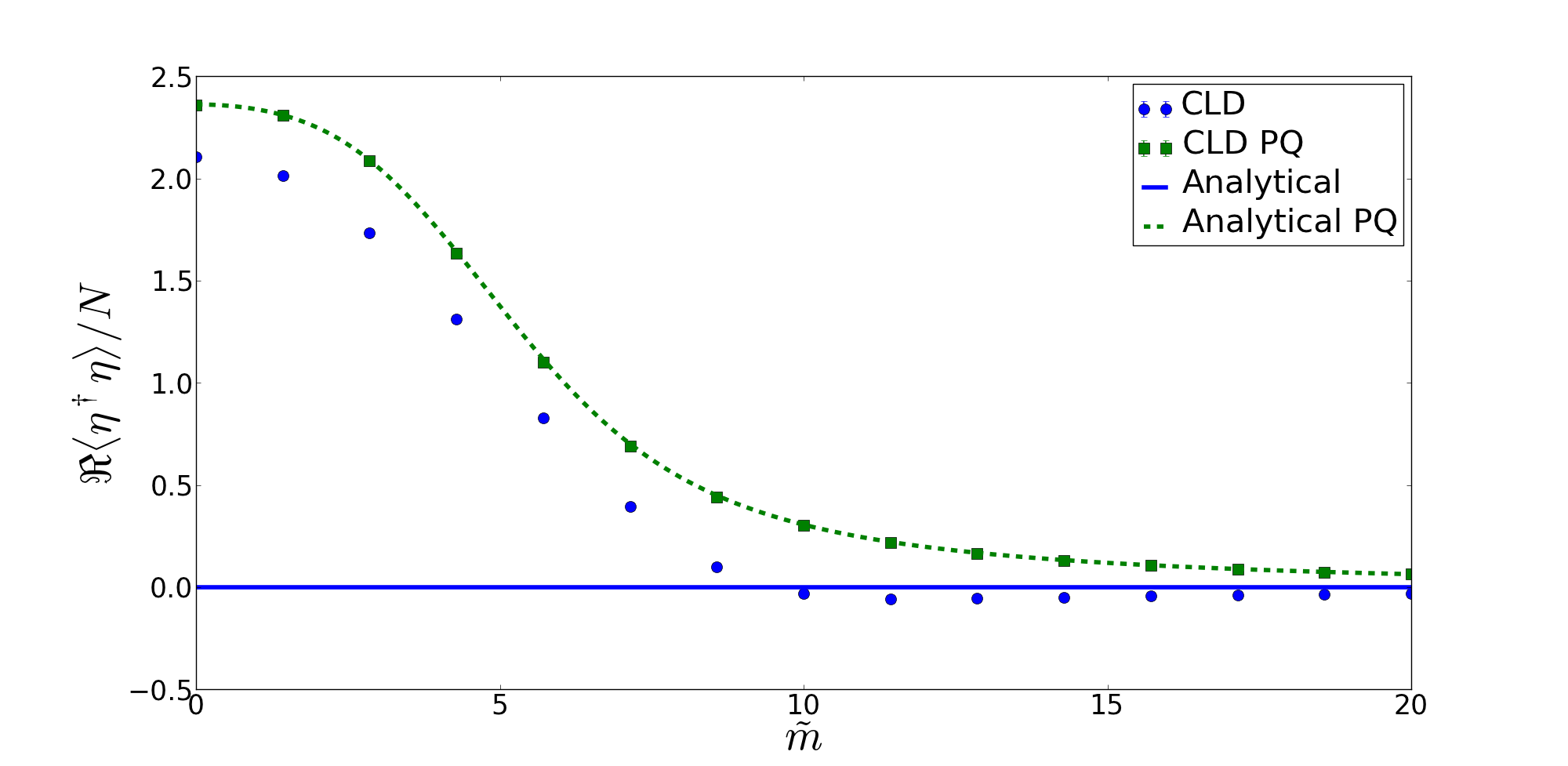}
\par\end{centering}

\caption{CLE measurement of the chiral condensate (top) and the baryon number
density (bottom) for $t=50$, $dt=10^{-4}$, $N=30$, $N_{f}=2$,
$\tilde{\mu}=2$ and a range of masses. The circular blue dots represent a
CLD simulation of the chiral random matrix theory in $\left(\ref{eq: RMT model 1}\right)$
and the blue line is the analytical result. We find correct convergence
for large masses, but failed convergence for small masses. The phase
quenched theory represented by the green squares and line, is simulated
correctly in the full mass range. Error bars based on statistics from
$3$ different simulations are present, but too small to see.\label{fig: condensate vs mass}}
\end{figure}
\begin{figure}
\begin{centering}
\includegraphics[width=14cm]{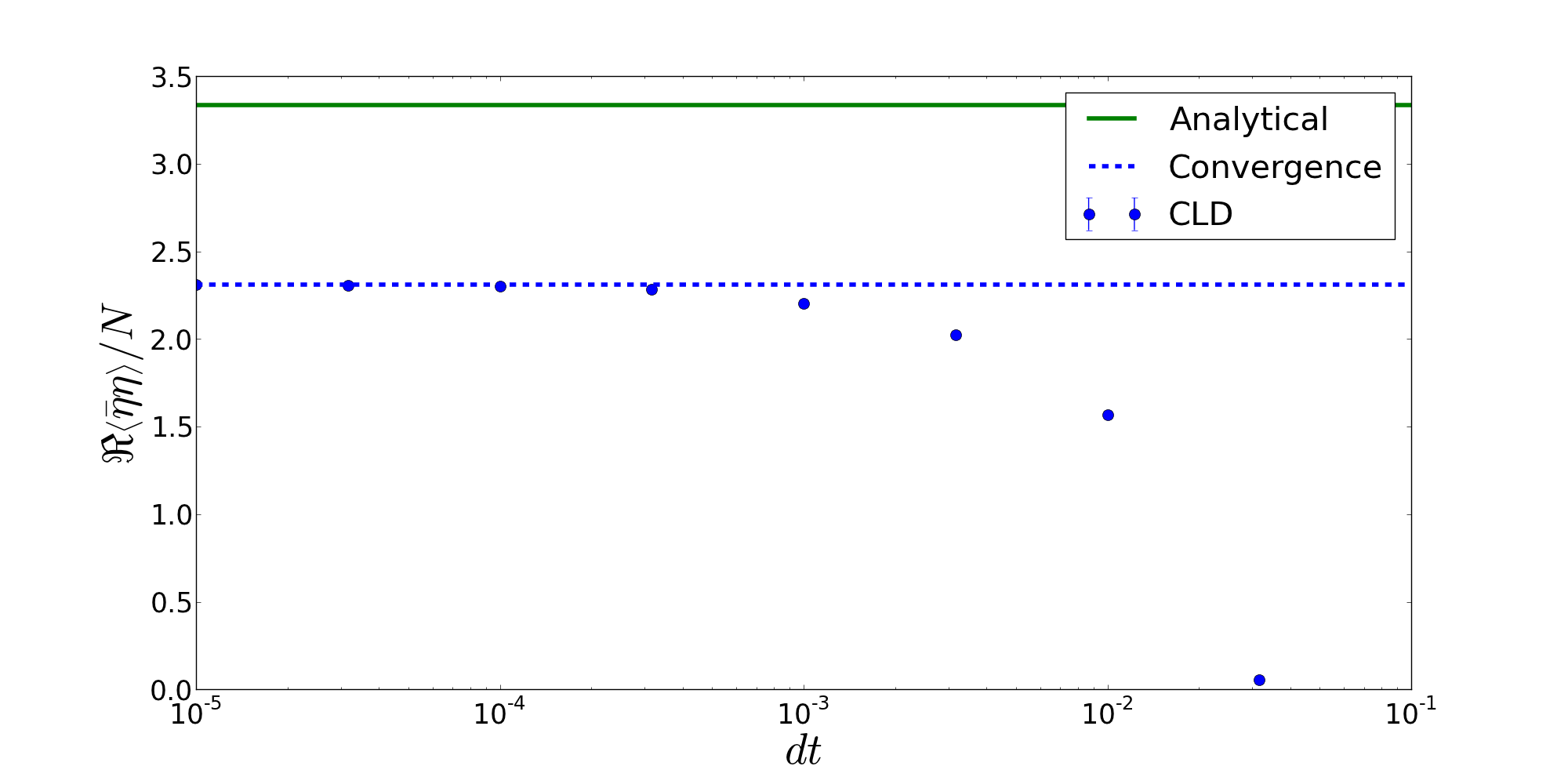}
\par\end{centering}

\caption{A plot of the chiral condensate as a function of $dt$ for $t=50$,
$N=30$, $N_{f}=2$, $\tilde{\mu}=2$ and $\tilde{m}=5$. We find
that the simulation settles around $dt=10^{-4}$, but not to the 
true analytically derived value given by the green line. \label{fig: stepsize}}
\end{figure}

\subsection{Analytical results}

We will consider the theory with two mass degenerate flavors, 
$N_{f}=2$, and perform measurements of the mass dependent chiral condensate 
\begin{equation}
\frac{1}{N}\left\langle \bar{\eta}\eta\right\rangle =\frac{1}{N}\partial_{m}\log\left(Z\right),
\end{equation}
and the baryon number density 
\begin{eqnarray}
\frac{1}{N}\left\langle \eta^{\dagger}\eta\right\rangle  & = & \frac{1}{N}\partial_{\mu}\log\left(Z\right),
\end{eqnarray}
where $\bar{\eta}$ and $\eta$ represent the quark fields. It
may be shown \cite{Bloch:2012bh} that the baryon number density 
vanishes 
\begin{equation}
\frac{1}{N}\left\langle \eta^{\dagger}\eta\right\rangle _{analytical}=0,
\end{equation}
while the chiral condensate may be found from the closed form of the
partition function in \cite{Osborn:2004rf} 
\begin{equation}
\frac{1}{N}\left\langle \bar{\eta}\eta\right\rangle _{analytical}=\frac{2m\left[L_{N}^{0}\left(x\right)L_{N-1}^{2}\left(x\right)-L_{N+1}^{0}\left(x\right)L_{N-2}^{2}\left(x\right)\right]}{L_{N}^{0}\left(x\right)L_{N}^{1}\left(x\right)-L_{N+1}^{0}\left(x\right)L_{N-1}^{1}\left(x\right)},
\end{equation}
with $L_{j}^{k}$ being the generalized Laguerre polynomials and $x\equiv-Nm^{2}$. 
(Note that the $\mu$-independence of the partition function in chiral random 
matrix theory corresponds to the $\mu$-independence of chiral perturbation 
theory. The measure in the chiral random matrix theory is of course strongly 
$\mu$-dependent, cf.~(\ref{eq: RMT model 1}).)

For comparison, we will also simulate the phase quenched theory,
in which $\mbox{det}^{2}(D(\mu)+m)$ is replaced by $\left|\det(D(\mu)+m)\right|^2$. The phase quenched chiral random matrix theory 
has been solved analytically in \cite{Akemann:2004dr}. The
chiral condensate is given as
\begin{equation}
\frac{1}{N}\left\langle \bar{\eta}\eta\right\rangle _{analytical}^{PQ}=4m\frac{\sum_{j=0}^{N}\cosh(2\mu)^{-2j}L_{j}^{0}(x)L_{j-1}^{1}(x)}{\sum_{j=0}^{N}\cosh(2\mu)^{-2j}L_{j}^{0}(x)^{2}},
\end{equation}
while the baryon number density is
\begin{equation}
\frac{1}{N}\left\langle \eta^{\dagger}\eta\right\rangle _{analytical}^{PQ}=\frac{4\tanh(2\mu)}{N}\left(\frac{\sum_{j=0}^{N}\left(N-j\right)\cdot\cosh(2\mu)^{-2j}L_{j}^{0}(x)^{2}}{\sum_{j=0}^{N}\cosh(2\mu)^{-2j}L_{j}^{0}(x)^{2}}\right).
\end{equation}
Since the phase quenched theory can be simulated with real Langevin dynamics
it serves as a partial check of the numerics.

\subsection{Langevin dynamics}

In order to optimize the simulation we reduce the dimensionality
of the argument of the determinant
\begin{eqnarray}
\det\left(D(\mu)+m\right) & = & \det\left(\begin{array}{cc}
m & X\\
Y & m
\end{array}\right) \ = \ \det\left(m^{2}-XY\right),\label{eq: expanding determinant}
\end{eqnarray}
where
\begin{equation}
X\equiv i\cosh(\mu)\Phi+\sinh(\mu)\Psi
\end{equation}
and
\begin{equation}
Y\equiv i\cosh(\mu)\Phi^{\dagger}+\sinh(\mu)\Psi^{\dagger}.
\end{equation}
Exponentiating the fermion determinant
\begin{equation}
Z_{N}^{N_{f}}(m)=\int d\Phi d\Psi\:\exp\left(N_{f}\cdot\mbox{Tr}\left[\log\left(m^{2}-XY\right)\right]-N\cdot\mbox{Tr}[\Psi^{\dagger}\Psi+\Phi^{\dagger}\Phi]\right),\label{eq: RMT model 2}
\end{equation}
we obtain the action
\begin{equation}
S=N\cdot\mbox{Tr}[\Psi^{\dagger}\Psi+\Phi^{\dagger}\Phi]-N_{f}\cdot\mbox{Tr}\left[\log\left(m^{2}-XY\right)\right].
\end{equation}
To derive the drift terms for $a_{ij},b_{ij},\alpha_{ij},\beta_{ij}$
we write out the action in terms of these. The Gaussian part reads
\begin{eqnarray}
\mbox{Tr}\left[\Psi^{\dagger}\Psi+\Phi^{\dagger}\Phi\right] & = & \mbox{Tr}\left[(\Psi^{\dagger})_{ij}\Psi_{jk}+(\Phi^{\dagger})_{ij}\Phi_{jk}\right]\nonumber \\
 & = & \Psi_{ji}^{*}\Psi_{ji}+\Phi_{ji}^{*}\Phi_{ji}\nonumber \\
 & = & (\alpha_{ji}-i\beta_{ji})(\alpha_{ji}+i\beta_{ji})+(a_{ji}-ib_{ji})(a_{ji}+ib_{ji})\nonumber \\
 & = & \alpha_{ji}^{2}+\beta_{ji}^{2}+a_{ji}^{2}+b_{ji}^{2},
\end{eqnarray}
while the non diagonal terms of the determinant take the form
\begin{equation}
X_{ij}=i\cosh\left(\mu\right)(a_{ij}+ib_{ij})+\sinh\left(\mu\right)(\alpha_{ij}+i\beta_{ij})\label{eq: X_ij}
\end{equation}
and
\begin{eqnarray}
Y_{ij} & = & i\cosh\left(\mu\right)\left((a+ib)^{\dagger}\right)_{ij}+\sinh\left(\mu\right)\left((\alpha+i\beta)^{\dagger}\right)_{ij}\nonumber \\
 & = & i\cosh\left(\mu\right)(a_{ji}-ib_{ji})+\sinh\left(\mu\right)(\alpha_{ji}-i\beta_{ji}).\label{eq: Y_ij}
\end{eqnarray}
We introduce the notation $G\equiv(m^{2}-XY)^{-1}$ and calculate
the drift term for $a_{mn}$ (ignoring the cut) 
\begin{eqnarray}
-\frac{\partial S}{\partial a_{mn}} & = & -2Na_{mn}-N_{f}\cdot\mbox{Tr}\left[\left((m^{2}-XY)^{-1}\right)_{li}\partial_{a_{mn}}\left(X_{ij}Y_{jk}\right)\right]\nonumber \\
 & = & -2Na_{mn}-N_{f}\cdot\mbox{Tr}\left[G_{li}\left(i\cosh\left(\mu\right)\delta_{mi}\delta_{nj}Y_{jk}+i\cosh\left(\mu\right)X_{ij}\delta_{mk}\delta_{nj}\right)\right]\nonumber \\
 & = & -2Na_{mn}-N_{f}i\cosh\left(\mu\right)\cdot\left[G{}_{ki}\left(\delta_{mi}\delta_{nj}Y_{jk}+X_{ij}\delta_{mk}\delta_{nj}\right)\right]\nonumber \\
 & = & -2Na_{mn}-N_{f}i\cosh\left(\mu\right)\cdot\left[G{}_{km}Y_{nk}+G_{mi}X_{in}\right]\nonumber \\
 & = & -2Na_{mn}-N_{f}i\cosh\left(\mu\right)\cdot\left[\left(\left(YG\right)^{\top}\right)_{mn}+\left(GX\right)_{mn}\right]\nonumber \\
 & = & -2Na_{mn}-iN_{f}\cosh(\mu)\left[R_{mn}+T_{mn}\right],\label{eq: drift 1}
\end{eqnarray}
where we have defined 
\begin{equation}
R_{mn}=\left(\left(YG\right)^{\top}\right)_{mn}
\end{equation}
and
\begin{equation}
T_{mn}=\left(GX\right)_{mn}.
\end{equation}
The derivation of the other drift terms follow the same logic
\begin{eqnarray}
-\frac{\partial S}{\partial b_{mn}} & = & -2Nb_{mn}+N_{f}\cosh\left(\mu\right)\cdot\left[R_{mn}-T_{mn}\right]\nonumber \\
-\frac{\partial S}{\partial\alpha_{mn}} & = & -2N\alpha_{mn}-N_{f}\sinh\left(\mu\right)\cdot\left[R_{mn}+T_{mn}\right]\nonumber \\
-\frac{\partial S}{\partial\beta_{mn}} & = & -2N\beta_{mn}-iN_{f}\sinh\left(\mu\right)\cdot\left[R_{mn}-T_{mn}\right].
\end{eqnarray}

Turning to the Langevin simulation we count $2\cdot2\cdot N^{2}$ real
degrees of freedom to be complexified
\begin{equation}
a_{ij},b_{ij},\alpha_{ij},\beta_{ij}\in\mathbb{R}\rightarrow a_{ij},b_{ij},\alpha_{ij},\beta_{ij}\in\mathbb{C},
\end{equation}
such that we effectively need to update $8N^{2}$ degrees of freedom
with each time step. 
Fast matrix manipulations may now be applied when updating the variables
according to the complex Langevin equation
\begin{equation}
u_{ij}\left(t+\Delta t\right)=u_{ij}\left(t\right)-\frac{\partial S}{\partial u_{ij}}\cdot\Delta t+dW_{ij},
\end{equation}
where $dW_{ij}$ is a Gaussian noise matrix 
\begin{equation}
\left\langle dW_{ij}\right\rangle =0,\quad\left\langle dW_{ij}\left(t\right)dW_{kl}\left(t'\right)\right\rangle =2dt\cdot\delta\left(t-t'\right)\delta_{ik}\delta_{jl},
\end{equation}
and $u_{ij}$ is representing any of $a_{ij},b_{ij},\alpha_{ij},\beta_{ij}$.

The chiral condensate is obtained 
by differentiating the partition function (\ref{eq: RMT model 2})
with respect to the mass
\begin{eqnarray}
\frac{\bar{\eta}\eta}{N} & = & \frac{N_{f}}{N}\partial_{m}\left(\mbox{Tr}\left[\log\left(m^{2}-XY\right)\right]\right)\nonumber \\
 & = & \frac{2mN_{f}}{N}\cdot\mbox{Tr}\left[\left(m^{2}-XY\right)^{-1}\right].\label{eq: chiral condensate variable}
\end{eqnarray}
Differentiating with respect to the chemical potential gives 
the baryon number density
\begin{eqnarray}
\frac{\eta^{\dagger}\eta}{N} & = & \frac{N_{f}}{N}\partial_{\mu}\left(\mbox{Tr}\left[\log\left(m^{2}-XY\right)\right]\right)\nonumber \\
 & = & \frac{N_{f}}{N}\left(\mbox{Tr}\left[\left(m^{2}-XY\right)^{-1}\partial_{\mu}\left(m^{2}-XY\right)\right]\right)\nonumber \\
 & = & \frac{N_{f}}{N}\left(\mbox{Tr}\left[\left(m^{2}-XY\right)^{-1}\left(-\left(\partial_{\mu}X\right)Y-X\left(\partial_{\mu}Y\right)\right)\right]\right).\label{eq: baryon number density variable}
\end{eqnarray}
\vspace{3mm}

To obtain the dynamics of the phase quenched theory, we use the 
relation in equation (\ref{eq: det*}) to write the absolute value 
as
\begin{eqnarray}
\left|\mbox{det}(D(\mu)+m)\right|^{N_{f}} & = & \left[\mbox{det}(D(\mu)+m)\mbox{det}^{\ast}(D(\mu)+m)\right]^{N_{f}/2}\nonumber \\
 & = & \mbox{det}^{N_{f}/2}(D(\mu)+m)\mbox{det}^{N_{f}/2}(D(-\mu)+m).
\end{eqnarray}
The dynamics may then be derived exactly as in the full theory.

\begin{figure}
\begin{centering}
\includegraphics[width=11.5cm]{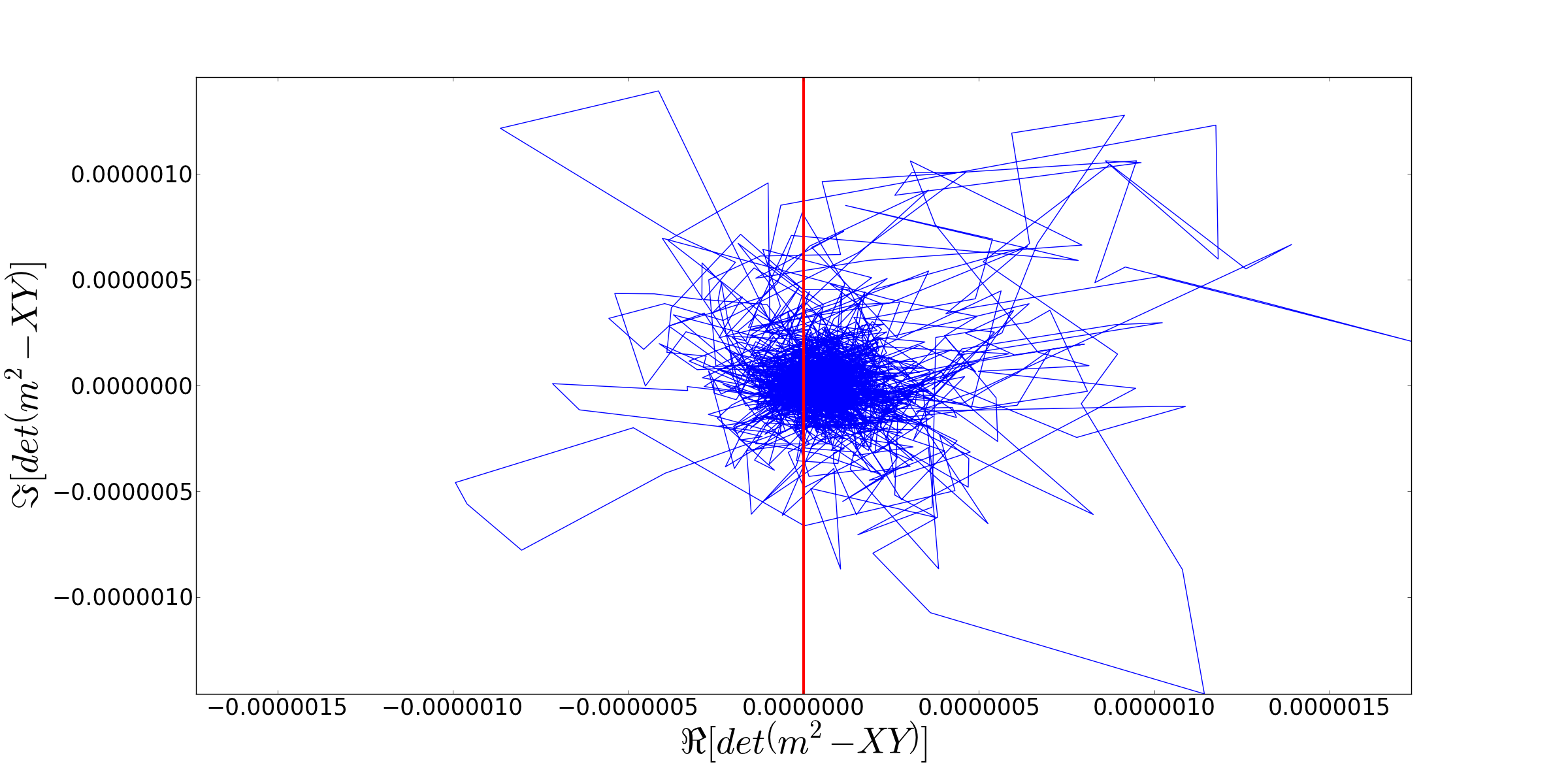}
\par\end{centering}

\begin{centering}
\includegraphics[width=11.5cm]{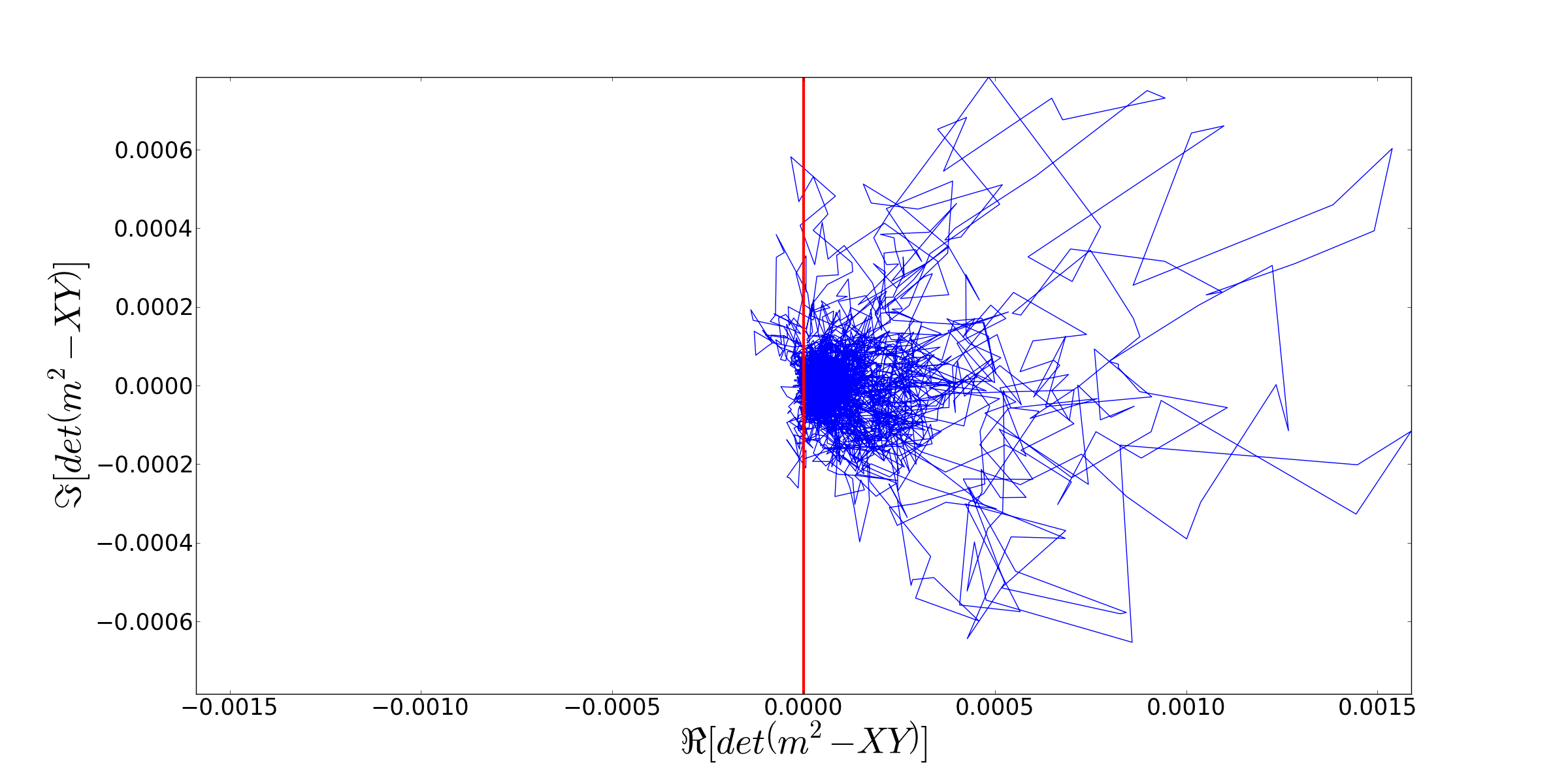}
\par\end{centering}

\begin{centering}
\includegraphics[width=11.5cm]{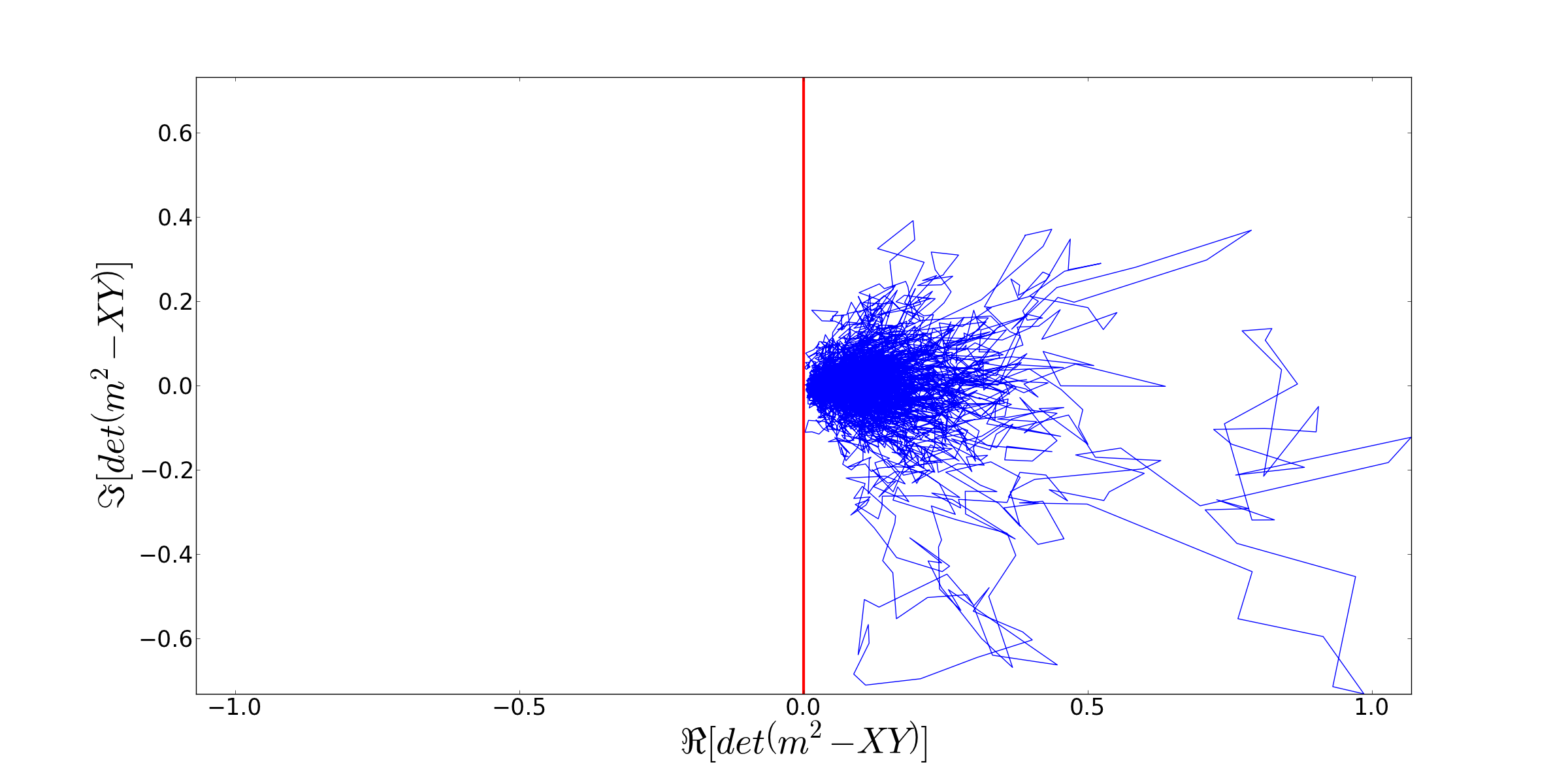}
\par\end{centering}

\caption{Flow plots of the determinant during a Langevin simulation of the
chiral random matrix theory for $T=50,$ $dt=10^{-4},$ $N=30$, $\tilde{\mu}=2$
and $\tilde{m}=5,10$, and respectively $15$. Comparing to the measurement 
of the chiral condensate and baryon number
density, we find that correct convergence, and restriction of the
determinant to the right half plane, share the same domain of masses.\label{fig:Scatterplots}}
\end{figure}

\begin{figure}
\begin{centering}
\includegraphics[width=11.5cm]{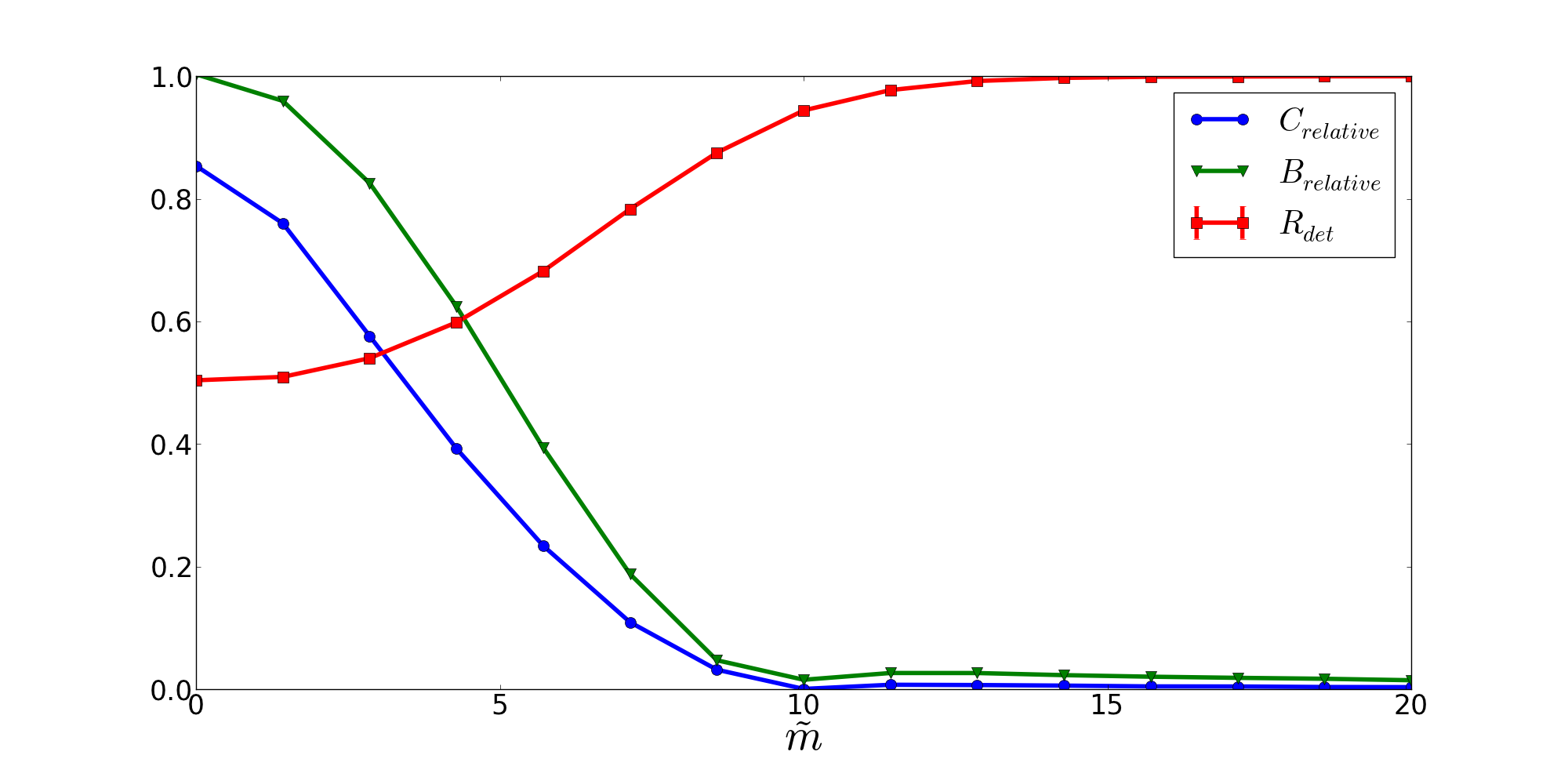}
\par\end{centering}

\begin{centering}
\includegraphics[width=11.5cm]{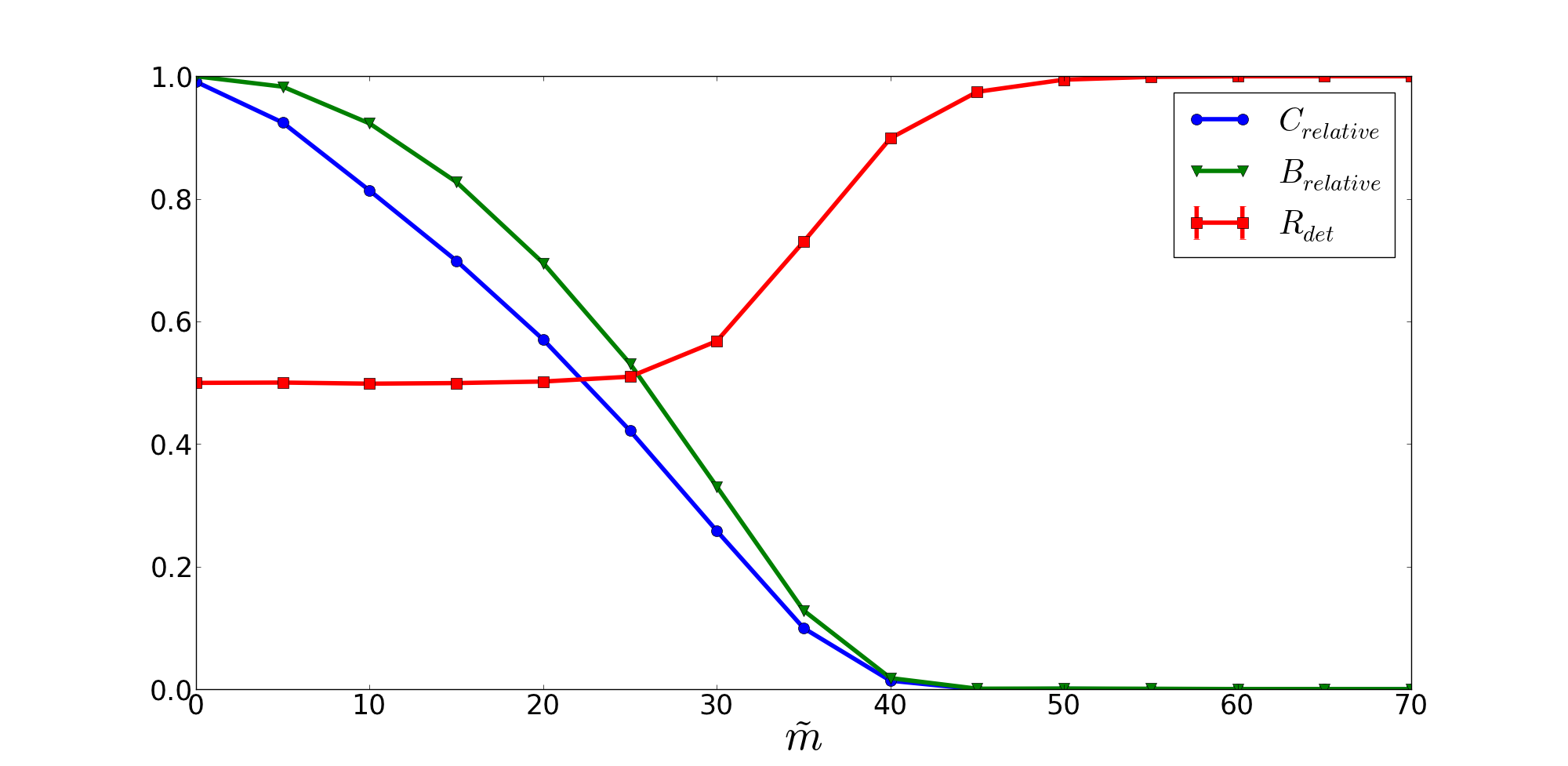}
\par\end{centering}

\begin{centering}
\includegraphics[width=11.5cm]{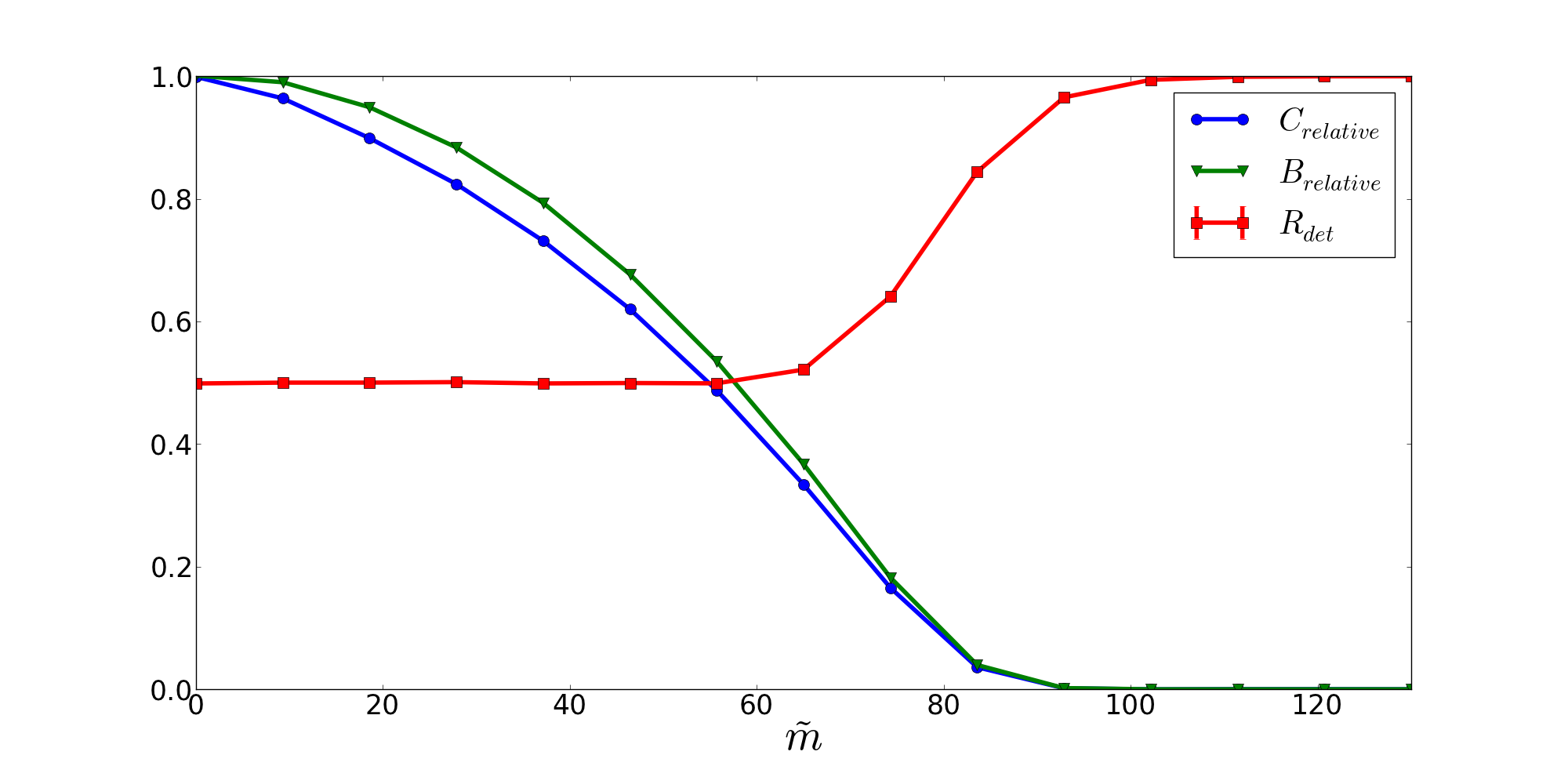}
\par\end{centering}

\caption{Comparison of the ratio of determinants in the positive half plane
and convergence of the the chiral condensate and baryon number
density during a Langevin simulation of the chiral random matrix theory for
$N=30$ and $\tilde\mu=2,5,8$. We observe that failed convergence and the
appearance of determinants in both half planes share the same domain
of masses. As expected the range varies substantially for the three 
values of $\tilde{\mu}$.\label{fig:ConvergenceVsPosRatio}}
\end{figure}

\subsection{Results}

Measurements of the chiral condensate are plotted in figure
\ref{fig: condensate vs mass} for $N=30$, $T=50$, $dt=10^{-4}$,
$N_{f}=2$, $\tilde{\mu}=\sqrt{N}\cdot\mu=2$ and a range of masses
$\tilde{m}=N\cdot m$.
 The blue disks in the plot represent the simulation, the blue line
is the analytical result, the green squares represent the phase quenched
simulation and the green line is the analytical result for the phase
quenched theory. The phase quenched theory converges correctly in
the full mass range, while the full theory only converges correctly
for masses greater than some critical value around, $\tilde{m}_{\mbox{critical}}\approx15$.
Measurements of the baryon number density have been plotted in 
the lower panel of figure
\ref{fig: condensate vs mass} for the same parameter values and same
representation of lines. Again the phase quenched theory converges
correctly in the full mass range, while the full theory fails to converge
for masses less than $\tilde{m}_{\mbox{critical}}$. The convergence
is admittedly not spot on for all $\tilde{m}>\tilde{m}_{\mbox{critical}}$
but it becomes better as $\tilde{m}$ is raised.

Since the drift becomes singular, when the determinant is close to zero,
the step size, $dt$, has to be chosen quite small in order to keep 
the finite step size effects at a minimum. In figure \ref{fig: stepsize} 
we plot $\left\langle \bar{\eta}\eta\right\rangle _{CLD}/N$
as a function of $dt$ for $t=50$, $N=30$, $N_{f}=2$, $\tilde{\mu}=2$
and $\tilde{m}=5$.
The result of the simulation is found to settle around $dt=10^{-4}$,
but not to the analytical result. Instead the dynamics fall closer
to the phase quenched theory, see figure \ref{fig: condensate vs mass}. 
We have simulated the chiral random 
matrix theory at various matrix sizes $N$ and the observed behavior 
is independent of $N$. In particular, the failed convergence at small 
masses is present even for $N=2$ where the singularity of the drift 
term is no problem to deal with numerically.

The failed convergence of the complex Langevin dynamics at small masses 
may be linked to the ambiguities of the drift term due to the logarithm by
plotting the values of the determinant during the simulation for different
values of the mass. This is done in figure \ref{fig:Scatterplots}
for $\tilde{\mu}=2$ and $\tilde{m}=5,10,15$; the cloud of determinants is found 
to move from the origin to the right half plane as $\tilde{m}$ is raised to
$\tilde{m}_{\mbox{critical}}$. 
This is exactly the point at which the cut of the logarithm 
safely may be ignored
and correct convergence sets in. In order to quantify this further
we plot in figure \ref{fig:ConvergenceVsPosRatio} the ratio of determinants
in the right half plane, $R_{\mbox{det}}$, with a measure of convergence
for the chiral condensate
\begin{equation}
C_{\mbox{relative}}\equiv\left|\left\langle \bar{\eta}\eta\right\rangle _{\mbox{CLD}}-\left\langle \bar{\eta}\eta\right\rangle _{\mbox{analytical}}\right|/\left|\left\langle \bar{\eta}\eta\right\rangle _{\mbox{analytical}}\right|,
\end{equation}
and the baryon number density
\begin{equation}
B_{\mbox{relative}}\equiv\left|\left\langle \eta^{\dagger}\eta\right\rangle _{\mbox{CLD}}-\left\langle \eta^{\dagger}\eta\right\rangle _{\mbox{analytical}}\right|/\max\left|\left\langle \eta^{\dagger}\eta\right\rangle _{\mbox{CLD}}\right|,
\end{equation}
against $\tilde{m}$ for three different values of $\tilde{\mu}$.
 The value of $\tilde{m}_{\mbox{critical}}$ changes as $\tilde{\mu}$
is raised, but the region of failed convergence corresponds to 
the presence of determinants in the left half plane in all three cases.
\vspace{0.15cm}

The complex Langevin simulation of chiral random matrix theory suggests 
a general criterion for actions containing a logarithm of a 
determinant, which must be satisfied in order for complex Langevin dynamics 
(using the standard derivative of the logarithm) to yield correct 
convergence: \textit{measurements
can only be trusted if the flow of the determinant does not frequently trace
out a path surrounding the origin.}

\section{Two $U\left(1\right)$ models}

\begin{figure}
\begin{centering}
\includegraphics[width=13cm]{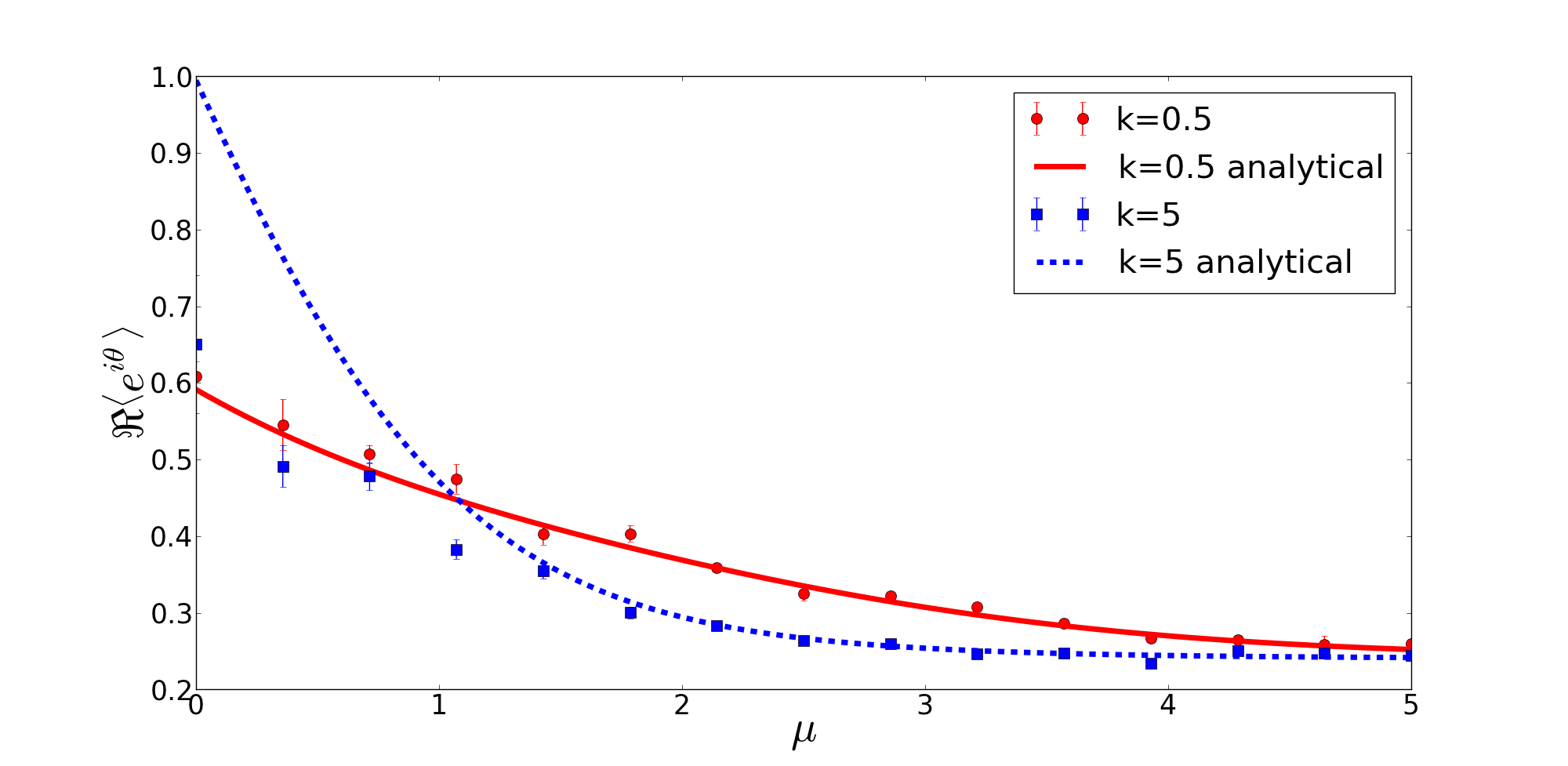}
\par\end{centering}

\caption{\label{fig: Aarts U1 measurements}Measurements of $\left\langle e^{i\theta}\right\rangle $
for $T=100$, $dt=10^{-4}$, $\beta=1$, $\mu\in\left[0,5\right]$
and the two values $\kappa=0.5$ and $\kappa=5$. For the first value
of $\kappa$ correct convergence is found in the full $\mu$ range,
while failed convergence is found for $\mu\lesssim1.5$ for the second
value of $\kappa$.}
\end{figure}

\begin{figure}
\includegraphics[width=14cm]{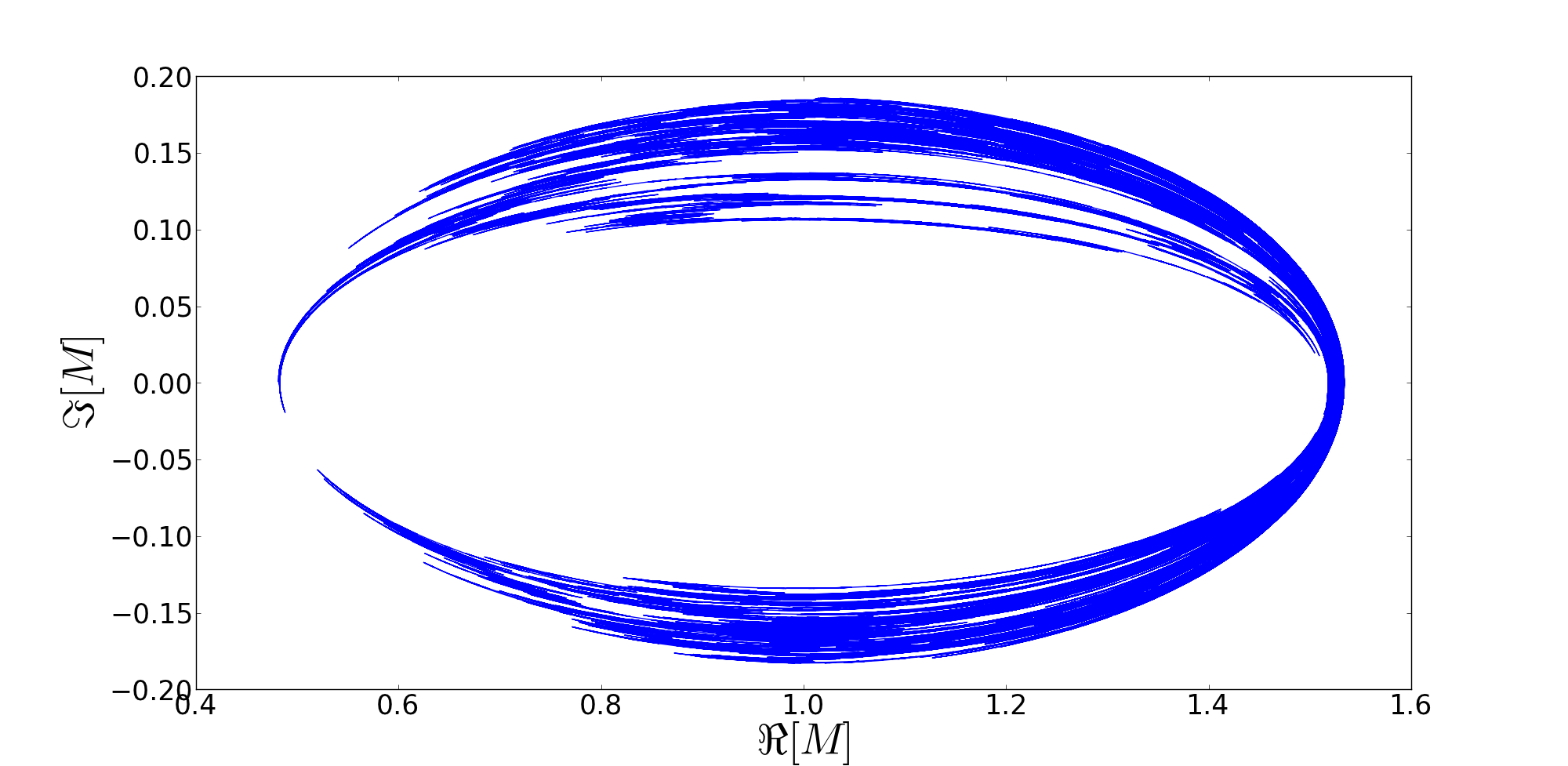}
\includegraphics[width=14cm]{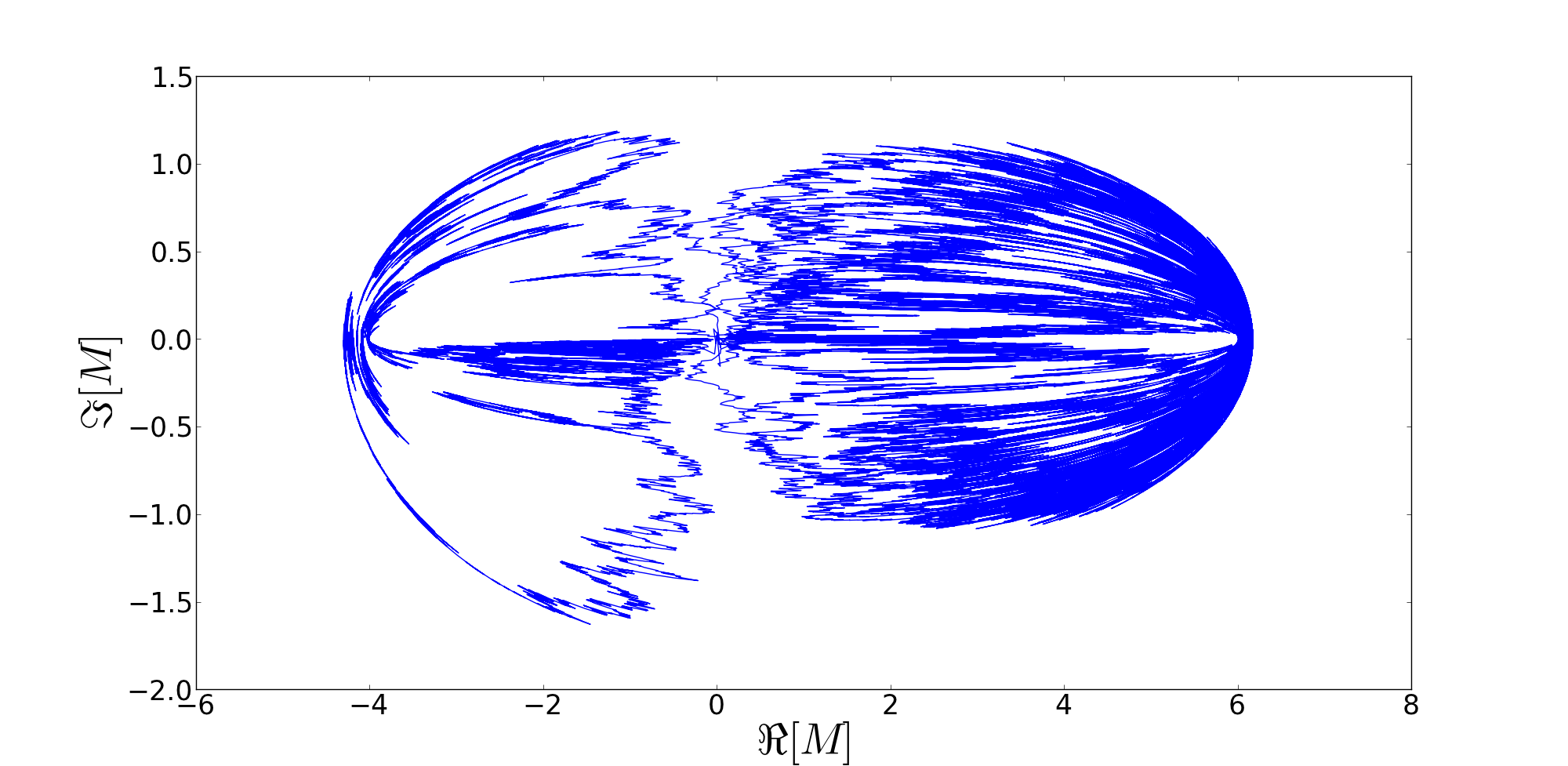}
\caption{\label{fig: M flow}Flow of the argument of the logarithm, 
$M\equiv1+\kappa\cos\left(\theta-i\mu\right)$,
for the U(1) link model in (\ref{eq: Aarts U1 action}) with $\mu=0.5$. 
Successful convergence is found for the path in the top figure with 
$\kappa=0.5$, where the cut safely may be ignored, while incorrect 
convergence is found for the path in the lower plot generated at $\kappa=5$. 
(Note that the ranges on the axis in the two plots are different.)}
\end{figure}

We shall now use the above criterion to understand the regions of 
successful and failed convergence in two previously studied 
$U\left(1\right)$ models \cite{Aarts:2008rr,Aarts:2010gr}. The action 
in the first model is
\begin{equation}
S=-\beta\cos\theta-\log\left(1+\kappa\cos\left(\theta-i\mu\right)\right),\label{eq: Aarts U1 action}
\end{equation}
and leads to the drift term (again ignoring the cut)
\begin{equation}
-\frac{\partial S\left(\theta\right)}{\partial\theta}=-\beta\sin\theta-\frac{\kappa\sin\left(\theta-i\mu\right)}{1+\kappa\cos\left(\theta-i\mu\right)}.
\end{equation}
In \cite{Aarts:2008rr} successful convergence of complex Langevin with the 
above drift term was found for values of $\kappa$ between 
$0$ and $1$. 
In figure \ref{fig: Aarts U1 measurements} we show plots of 
$\Re\langle e^{i\theta}\rangle$ 
with $\kappa=0.5$ and $\kappa=5$. As in \cite{Aarts:2008rr} we find 
perfect convergence for complex Langevin for $\kappa$ between 
$0$ and $1$ but for $\kappa=5$ we find failed convergence when the 
standard derivative of the logarithm is used.
The correct and failed convergence may be predicted by studying
the dynamics of the argument of the logarithm. As an example we plot in 
figure \ref{fig: M flow} two paths which $M$ traces out for $\mu=0.5$: 
With $\kappa=0.5$ the cut can safely be ignored, while for $\kappa=5$, 
where the convergence fails, the trajectory circles the origin.  
\vspace{3mm}

\begin{figure}
\begin{centering}
\includegraphics[width=13cm]{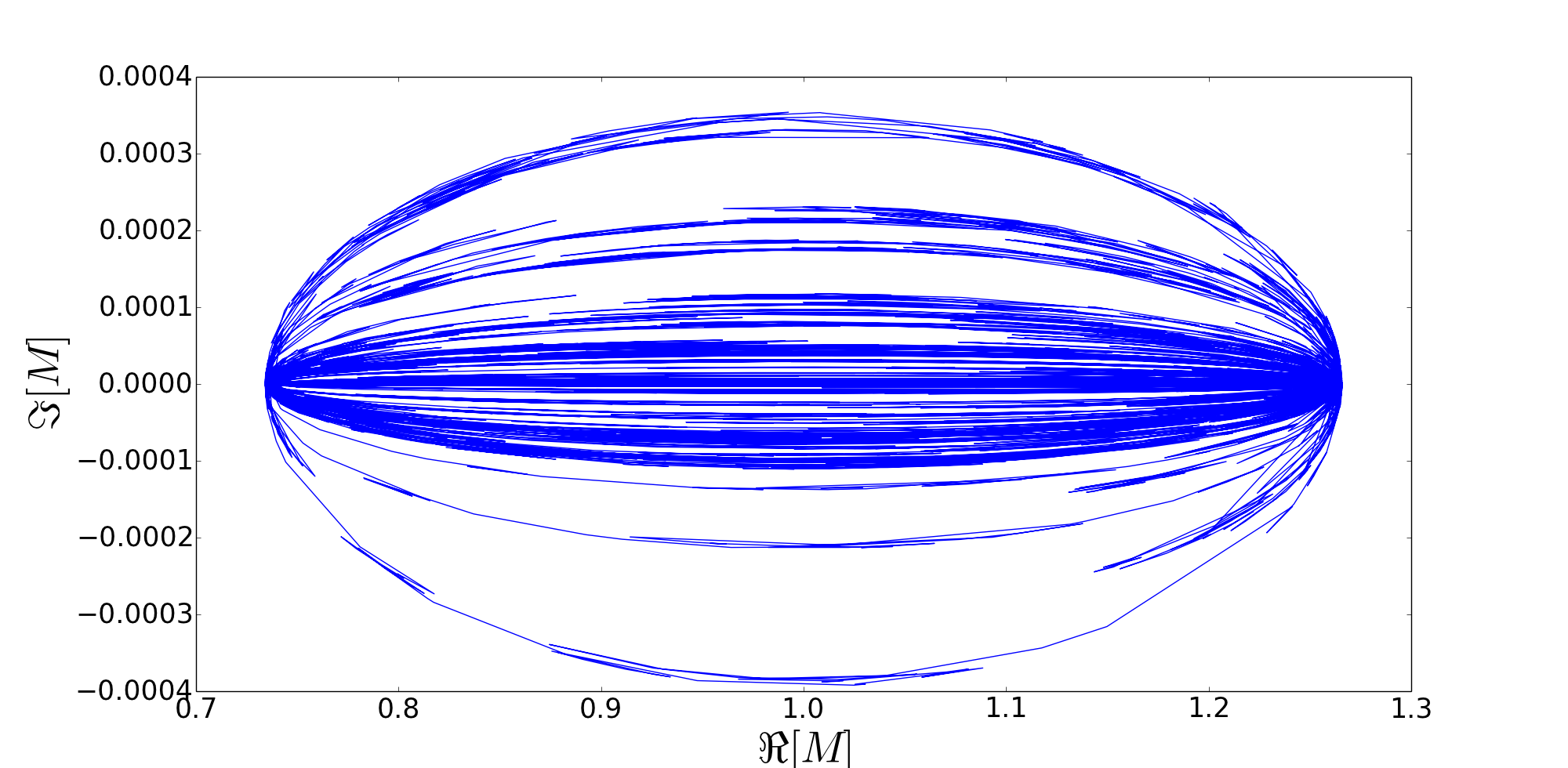}
\par\end{centering}

\caption{\label{fig:1dQCD} The trajectory which the argument of the 
logarithm in the action of a one dimensional QCD eigenvalue integral, 
Eq.~(\ref{1dQCD-M}), traces out during the complex Langevin simulation. 
The parameters used here are $\mu=1$, $\mu_c=0.4$, $n=4$ which correspond 
to the region with a strong sign problem in the original integral.  
The flow does not circle the origin and the complex Langevin simulation
converges nicely.}
\end{figure}

The second $U(1)$ model we consider, illustrates that an oscillating phase 
of the argument of the logarithm, 
when sampling the original real valued manifold of the integral, does 
not necessarily imply an oscillating phase, when sampling in the 
complexified space of the Langevin simulation. This example is 
found in the eigenvalue representation of one dimensional QCD 
first studied with complex Langevin in \cite{Aarts:2010gr}. The partition 
function is given by
\begin{eqnarray}
Z & = & \int_{-\pi}^{\pi}\frac{d\alpha}{2\pi} \, e^{\log\left(M\left(\alpha\right)\right)},\label{eq:1D-QCD-part-func}
\end{eqnarray}
with the argument of the logarithm given by
\begin{equation}
M\left(\alpha\right)=1-\frac{\cosh\left[n\left(\mu+i\alpha\right)\right]}{\cosh\left(n\mu_{c}\right)}.
\label{1dQCD-M}
\end{equation}
As demonstrated in \cite{Aarts:2010gr} complex Langevin successfully 
computes the average of the chiral condensate
\begin{equation}
\Sigma(\alpha) =\frac{1}{\sinh(\mu+i\alpha)+\sinh(\mu)}
\end{equation}
even for large values of $n$ and $\mu>\mu_c$ where $M$ oscillates wildly 
in the original real valued integral. At first this may
appear to be in contrast to what we have observed in chiral random matrix 
theory. In particular, since the oscillations are essential in order 
to obtain the correct chiral condensate \cite{Ravagli:2007rw} when working 
with a real valued angle $\alpha$.
However, as shown in figure \ref{fig:1dQCD} the path which 
$M\left(\alpha\right)$ traces out during the complex Langevin 
simulation, surrounds $M=1$ and does not enter the negative half plane. 
Therefore successful measurements may be performed even when one ignores the 
ambiguities related to the logarithm.

\section{Conclusions and outlook}

We have performed complex Langevin simulations of a chiral random matrix 
theory which, in the microscopic limit, is equivalent to QCD at nonzero 
chemical potential. The virtue of the random matrix theory is that it 
allows for an exact analytic solution and hence to test if
the complex Langevin dynamics converges correctly. 
While the complex Langevin simulation works at larger 
values of the quark mass we found that it converges incorrectly  
for small masses. As in full QCD the sign problem 
in chiral random matrix theory is due to the complex 
valued fermion determinant. In the complex Langevin dynamics 
the fermion determinant enters through a logarithmic term in the action. 
Because the logarithm is a multivalued function it leads to an ambiguity
when defining the Langevin force: should one take the derivative of the 
logarithm using the standard form, or should one try to incorporate 
the cut of the principal part of the logarithm. The standard choice used in 
the literature is to ignore the cut of the principal part of the logarithm 
and this is also the choice with which we implemented the complex Langevin 
dynamics for chiral random matrix theory.
We have demonstrated that the failure of the complex Langevin dynamics 
at small values of the quark mass occurs in the region of parameters,  
where the fermion determinant frequently circles the origin during the 
complex Langevin flow. This is exactly the region where the ambiguity of 
the logarithm is relevant.  
Based hereon it is natural to propose a criterion, which must be 
satisfied in order to safely use the standard form of the derivative of 
the logarithm in complex Langevin dynamics; the determinant is not allowed
frequently to trace out a path surrounding the origin. This criterion has 
been used to predict regions of successful and failed convergence in 
two previously studied $U\left(1\right)$ models. 

The phase of the fermion determinant in QCD at nonzero chemical potential 
is known to cover the full 
range $\left[-\pi,\pi\right]$, when sampling on the real manifold, see 
eg.~\cite{LSV,GMS}. As we have exemplified in a $U(1)$ model this behavior 
is not necessarily reproduced
in the complexified Langevin space, so it is possible that the cut
may be ignored in a simulation of QCD. This may of course be checked
explicitly in any given complex Langevin simulation of full QCD by 
plotting the trajectory of the fermion determinant during the Langevin flow.

The success and failure, observed as a function of the quark mass in 
the Langevin simulation of chiral random matrix theory, is quite similar to
that observed in the 0+1 dimensional Thirring model \cite{Pawlowski:2013pje}.
Langevin simulations of the Thirring model have not been performed as a part
of this work, but it is natural to expect that the same mechanism is 
responsible for the incorrect dynamics observed in \cite{Pawlowski:2013pje}.

Other issues of complex Langevin dynamics have been solved by introduction 
of techniques
such as the adaptive step size and gauge cooling. It is possible that
the ambiguities associated with the logarithm of the fermion determinant 
may be avoided by altering the dynamics of the simulation. This
is the challenge ahead\footnote{Some first attempts in this direction 
have been presented in \cite{AMspeciale}}. 
\vspace{3mm}

\noindent
{\bf Acknowledgments:} We wish to thank Poul Henrik Damgaard, Gert Aarts, 
Joyce C.~Meyers, as well as  
participants and organizers of XQCD 2013 at the AEI in Bern for discussions. 
Jac Verbaarschot and Gert Aarts are thanked for insightful comments on the 
manuscript. The work 
of KS was supported by the {\sl Sapere Aude program} of The Danish 
Council for Independent Research. 
 
\appendix

\section{$U(1)$ model with flow towards the real axis}
\label{app:A}

In this appendix we revisit the model introduced in 
\cite{Ambjorn:1986fz} as a simplification of a 
$U\left(1\right)$-link model. This case is special in that 
the argument of the logarithm flows to the real axis rather 
than fluctuating in the complex plane. 

\begin{figure}
\begin{centering}
\includegraphics[width=14cm]{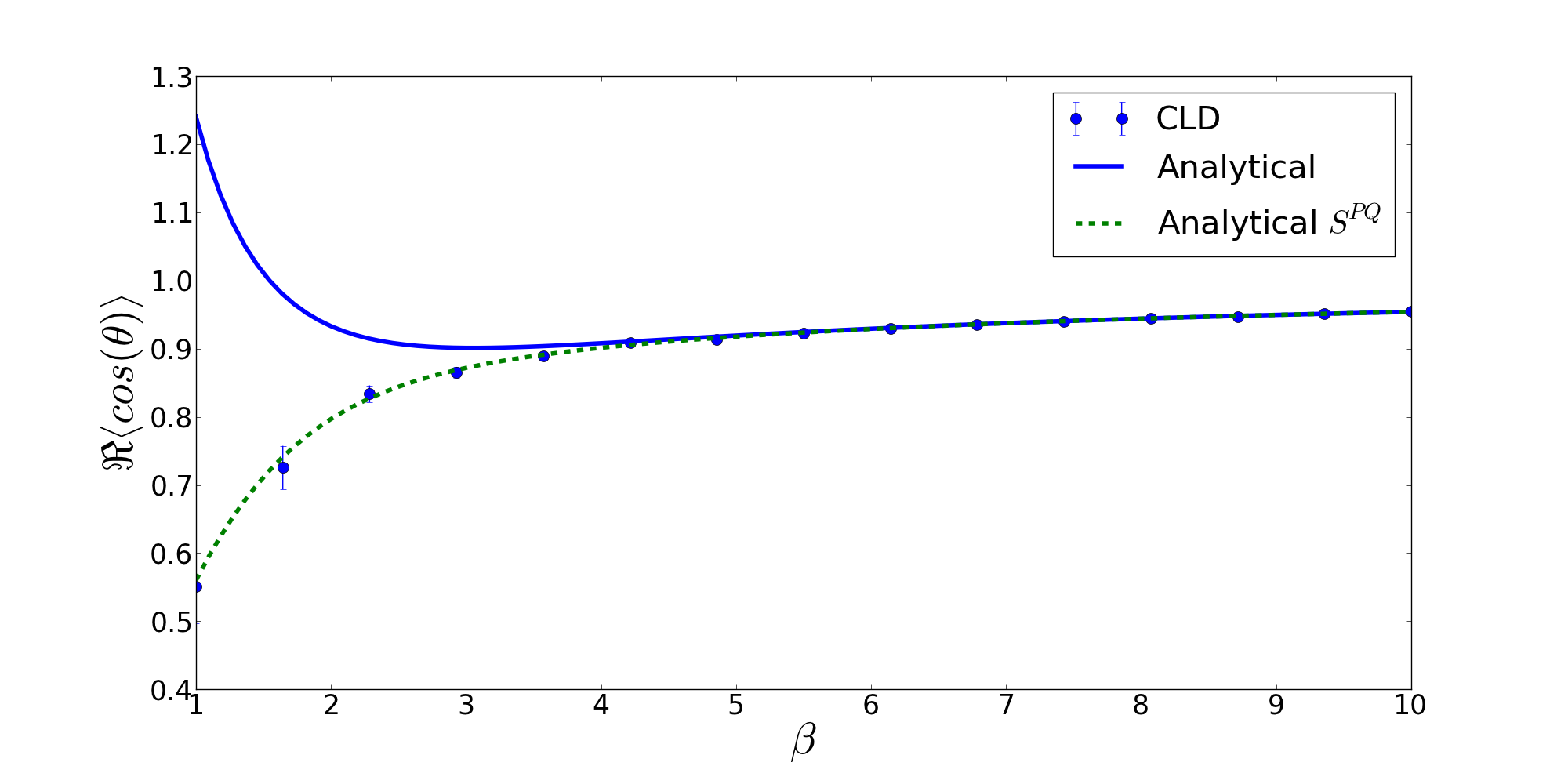}
\par\end{centering}

\caption{\label{fig:<cos>}Langevin measurement of $\cos\theta$ for $T=200$,
$dt=0.002$ and the drift term given in (\ref{eq: ambj=0000F8rn drift}).
The dynamics produces failed convergence for small $\beta$, where
the results of the phase quenched theory in (\ref{eq: real action Ambj=0000F8rn})
are produced instead.}
\end{figure}

\begin{figure}
\begin{centering}
\includegraphics[width=14cm]{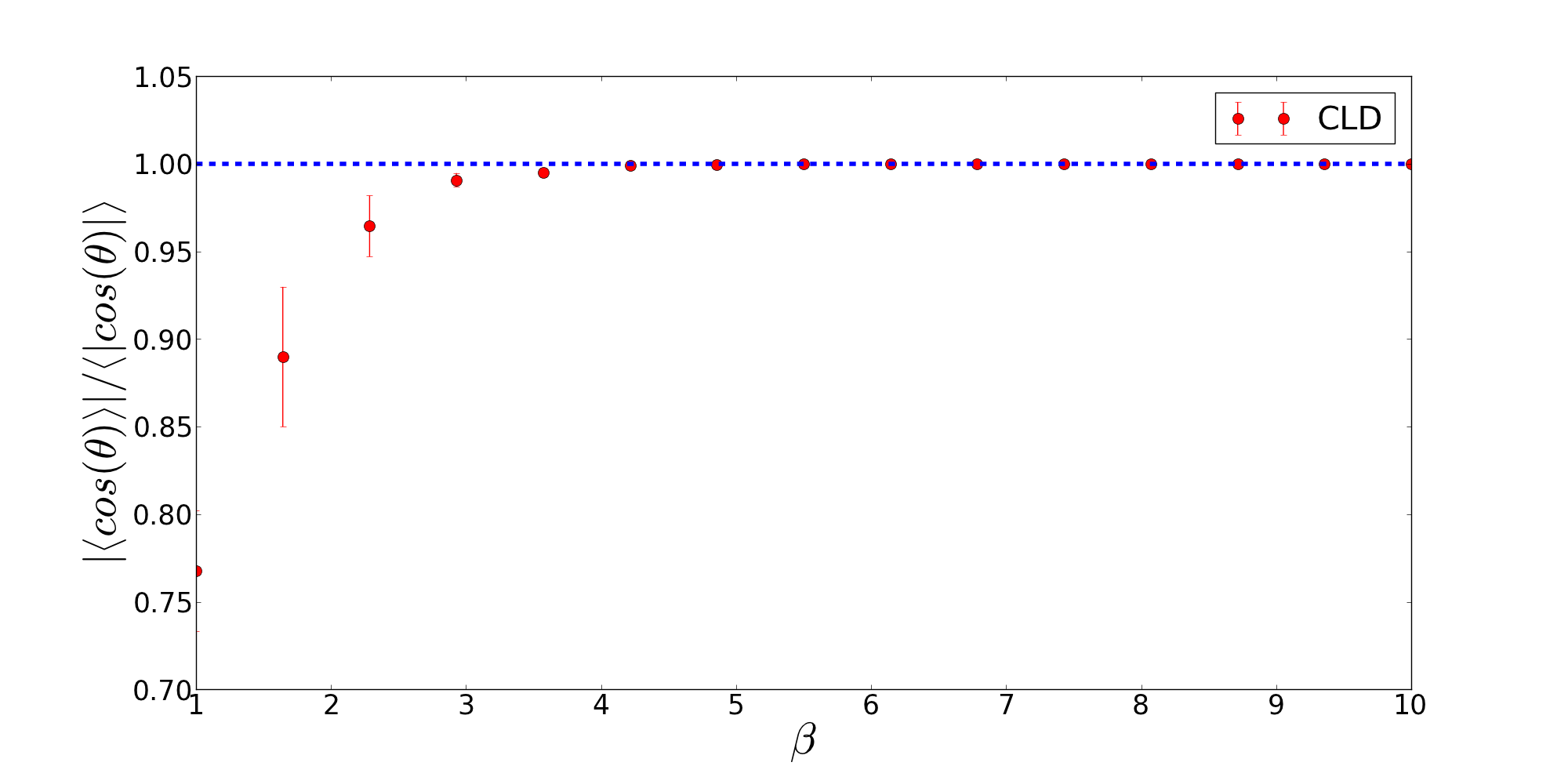}
\par\end{centering}

\caption{\label{fig:SignProblem-Ambj=0000F8rn}Plot of the simulated sign problem
$\left|\left\langle \cos\theta\right\rangle _{CLD}\right|/\left\langle \left|\cos\theta\right|\right\rangle _{CLD}$
for the drift in (\ref{eq: ambj=0000F8rn drift}) with $T=200$, $dt=0.002$
and a range of $\beta$ values. We find that the sign problem region
is coinciding with the region of failed convergence in figure \ref{fig:<cos>}.}
\end{figure}

Consider a single compact degree of freedom, 
$\theta\in\left[-\pi,\pi\right]$, and an action
\begin{equation}
S(\theta)=-\beta\cos\theta-\log\left(\cos\left(\theta\right)\right).
\end{equation} 
Calculating the drift term (by ignoring the cut) 
\begin{equation}
-\frac{\partial S\left(\theta\right)}{\partial\theta}=-\beta\sin\theta-\tan\theta\label{eq: ambj=0000F8rn drift}
\end{equation}
it was found \cite{Ambjorn:1986fz} that measurements of 
$\left\langle \cos\theta\right\rangle$ are successful for large values 
of $\beta$ but fail for $\beta\lesssim3.5$, see figure \ref{fig:<cos>}. 
 The region of failed convergence, $\beta\lesssim3.5$, may be predicted
by plotting a measure of the simulated sign problem, $\left|\left\langle \cos\theta\right\rangle _{CLD}\right|/\left\langle \left|\cos\theta\right|\right\rangle _{CLD}$,
as done in figure \ref{fig:SignProblem-Ambj=0000F8rn}. When $\beta$
is lowered to around $3.5$, the argument of the logarithm, $\cos\theta$,
starts changing sign, and this is also the value, below which the measurements
with the complex Langevin dynamics fail.

\begin{figure}
\begin{centering}
\includegraphics[width=14cm]{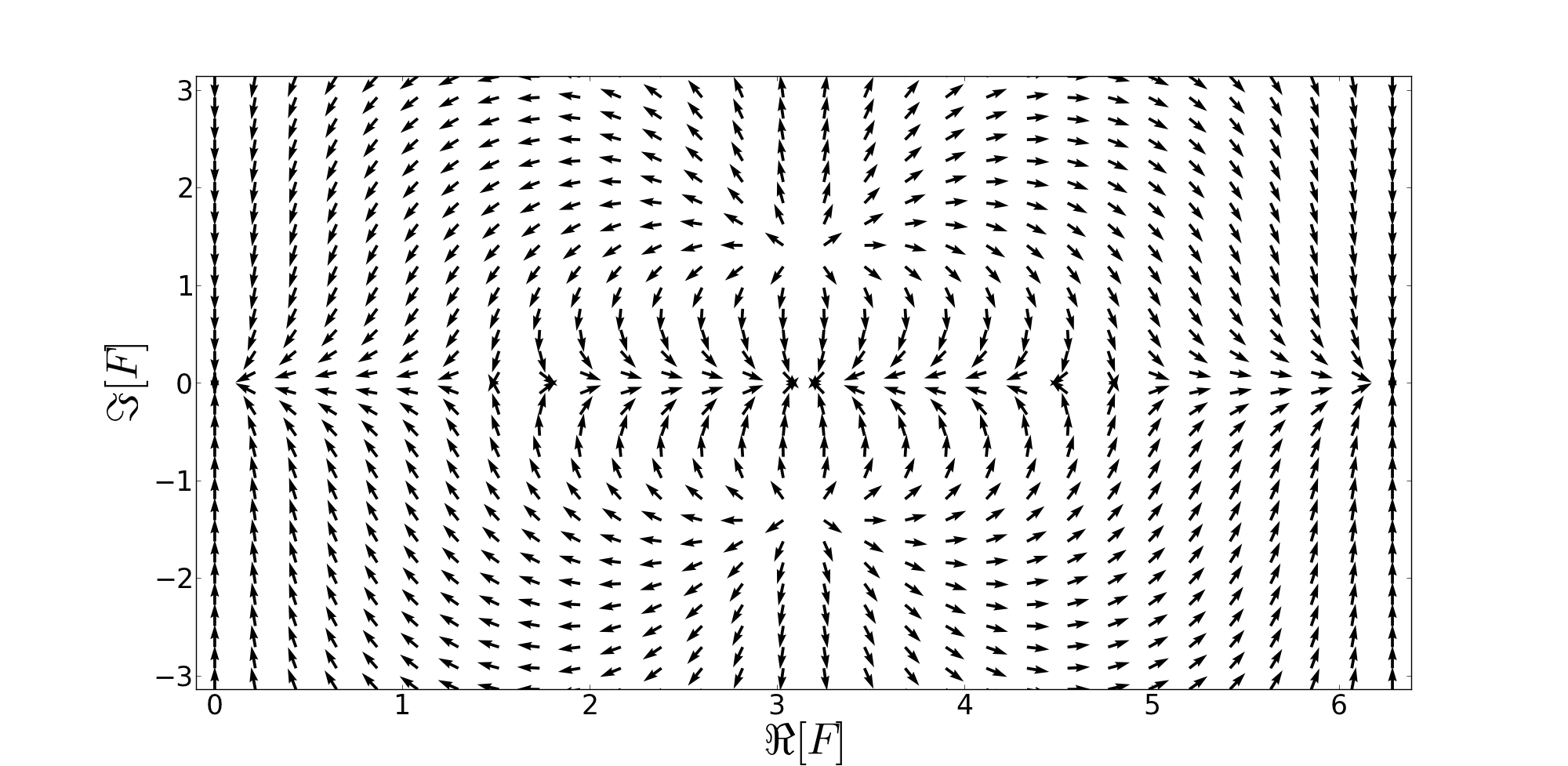}
\par\end{centering}
\caption{\label{fig:Flow-diagram-Ambj=0000F8rn}Flow diagram for the drift
in (\ref{eq: ambj=0000F8rn drift}) with $\beta=0.5$. The trajectories are attracted to the real axis.}
\end{figure}

Also shown in figure \ref{fig:<cos>} is the result for 
$\left\langle \cos\theta\right\rangle$ computed with the phase quenched action
\begin{equation}
S^{\rm PQ}(\theta)=-\beta\cos\theta-\Re\left[\log\left(\cos\left(\theta\right)\right)\right].\label{eq: real action Ambj=0000F8rn}
\end{equation}
As already observed in \cite{Fujimura:1993cq} we see that the measurement 
converges to the phase quenched results. The reason why the convergence to the 
phase quenched result is exact in this case can be understood from  
the flow diagram in figure \ref{fig:Flow-diagram-Ambj=0000F8rn}.
The flow is attracted to the real axis, where the logarithm takes
the form
\begin{equation}
\log x=\Re\left[\log x\right]+i\pi\cdot\Theta\left(-x\right),
\end{equation}
so ignoring the discontinuity of the imaginary part, is in this case 
tantamount to working in the phase quenched theory. In   
\cite{Fujimura:1993cq} attempts to include the effect of the step function 
in the complex Langevin flow where discussed. These have been reexamined 
in \cite{AMspeciale}.

It should be noticed that the singular behavior of the logarithm at
the origin does not seem to pose a problem. One may convince oneself of
this by studying the action
\begin{equation}
S(x)=ax^{2}-\log(x^{2}),
\end{equation}
which forces $x$, and therefore also $x^{2}$, close to the singularity
for $a>0$. Simulations works beautifully for adaptive or small step
sizes, but if the argument of the logarithm is allowed to change phase
as for the slightly generalized action
\begin{equation}
S(x)=x^{2}-\log(m^{2}-x^{2}),
\end{equation}
then complex Langevin dynamics (using $\partial_{z}\log z=1/z$) fails.

\end{document}